\documentclass{elsart1p}
\usepackage{graphicx}
\usepackage{amsmath}
\usepackage{amssymb}

\begin{document}

\begin{frontmatter}

\title{Equilibration of two non-extensive subsystems in a parton cascade model}

\author{Tam\'as S. B\'ir\'o and}
\author{G\'abor Purcsel}
\ead{purcsel@rmki.kfki.hu}

\address{KFKI Research Institute for Particle and Nuclear Physics\\
of the Hungarian Academy of Sciences\\
H-1525 Budapest, P.O.Box 49, Hungary}

\begin{abstract}
We study the process of equilibration between two non-extensive subsystems in the framework of a particular non-extensive Boltzmann equation. We have found that even subsystems with different non-extensive properties achieve a common equilibrium distribution.
\end{abstract}

\begin{keyword}
non-extensive \sep statistics \sep Boltzmann equation

\PACS 05.70.Ln
 \sep 05.20. Dd
\end{keyword}

\end{frontmatter}


\section{Introduction}\label{intro}

    Power-law tailed spectra occur in a wide range of physical phenomena,
re-association in folded proteins \cite{protein},
fluxes of cosmic rays \cite{cosmicray},
turbulence \cite{turb1},
finance and economics \cite{finance1},
electron-positron annihilation \cite{epannih},
motion of Hydra cells \cite{hydra},
epilepsy \cite{epilepsy},
linguistics \cite{ling},
nuclear physics \cite{nuclphys},
astrophysics \cite{astro},
field theories and cosmology \cite{Beck:2003wj},
scientific citations \cite{cit1},
distributions of individual succes of musicians \cite{music1},
urban agglomerations \cite{urban},
internet phenomena \cite{internet},
phase transformations \cite{Kayacan:2005yj},
in algorithms for global optimization (e.g., the generalized simulated annealing) and related computational methods \cite{opt01} and
information theory \cite{Jizba:2005nn}.

We are particularly interested in heavy ion collisions where cut power-law distributions describe transverse momentum (\(p_T\)) spectra at low and intermediate values \cite{TSBJPG}. The conventional approach \cite{Bass:2004di} to the distribution function fits a Boltzmann-Gibbs exponential characterized by a single temperature and the deviation from the exponential at the tail is attained to non-equilibrium effects. A transverse expansion distorts the locally thermal spectra; at low \(p_T\) by suppression, at high \(p_T\) by a blue shift factor in the temperature.
    
    Some further attempts have been made to interpret power-law tailed spectra as equilibrium phenomena for the whole \(p_T\) range: Non-extensive thermodynamicses predict such distributions \cite{nuclphys,Tsallis1}. It has been shown, that dynamicses among (quasi-) particles can be constructed, leading to power-law tailed stationary distributions \cite{TSBPGPRL}. This parton cascade model, referred to as non-extensive Boltzmann equation (NEBE), generates stationary states subject to non-extensive kinetic energy addition rules. We have shown that an infinite class of equilibrium distributions can be established depending on the way we generalize the Boltzmann equation \cite{TSBGK}. In this paper we point out that cut power-law distributions play a special role among all possible functions: The rule applied in the non-extensive Boltzmann equation leading to such distributions is the leading term in a low-energy expansion of a general associative rule.
    
    A crucial question towards non-extensive thermodynamics has been discussed in \cite{Nauenberg1}, namely how do two non-extensive thermodynamical systems thermalize, if at all. In this article we aim to answer this question by performing momentum-space simulations within the NEBE model.


\section{Non-Extensive Boltzmann Equation (NEBE)}

	In the non-extensive extension of the Boltzmann equation we keep the original factorizing form for the two-body distributions,
	\begin{equation}
		\frac{\partial}{\partial t} f_1 = \int_{234} w_{1234} \left( f_3f_4-f_1f_2 \right),
	\end{equation}
	but in the transition rate
	\begin{equation}
		w_{1234} = M_{1234}^2\cdot \delta\left(\left(\vec p_1+\vec p_2\right)-\left(\vec p_3+\vec p_4\right)\right) \cdot \delta\left(h\left(E_1,E_2\right)-h\left(E_3,E_4\right)\right),
	\end{equation}
	the two-body energy composition is generalized to a rule \(h\left(E_1,E_2\right)\), which is not necessarily the simple addition.
 There can be many physical sources for this deformation of the kinetic energy addition rule; the most general being a pair interaction due to a potential whose value differs before and after the two body collision. This way corrections occur, which (e.\ g.\ by using a virial theorem) may be expressible in terms of the individual kinetic energies of the colliding subsystems (particles).
 Here \(E_i=\sqrt{\vec p_i^2+m_i^2}\) are the free kinetic energies of relativistic particles with mass \(m_i\). In a heavy-ion collision they are regarded as the asymptotic energies detected after desintegration of the system.

	The energy composition rule \(h(E_1,E_2)\) contains contributions stemming from the pair interaction. It is not trivial whether these can always be divided to one-particle contributions, supporting a quasi-particle picture. In the followings we demonstrate that under quite general assumptions about the function \(h(x,y)\) the division of the total energy among free particles can be done.
	
	We assume that the generalized energy sum is associative,
	\begin{equation}
		h\left(h\left(x,y\right),z\right) = h\left(x,h\left(y,z\right)\right).   \label{assoc}
	\end{equation}
	Then due to a mathematical theorem \cite{mathbookFEiAS} a strict monotonic function \(X(h)\) maps the energy composition rule to additivity
	\begin{equation}
		X\left(h\right) = X\left(x\right) + X\left(y\right). \label{Xadd}
	\end{equation}
	This solution \eqref{Xadd} of the functional equation \eqref{assoc} is unique up to a constant multiplicative factor. 
	In this case the stationary solution of the non-extensive Boltzmann equation is given by 
	\begin{equation}
		f\left(\vec p\right) = \frac1Z\: e^{-X\left(E\right)/T},
	\end{equation}
	and \(X(E)\) is regarded as the energy of a quasi-particle.

	One finds several examples for the energy composition rule \(h(x,y)\). Some of them lead to thermodynamicses where the entropy formula or the distribution function is familiar: e.g.\ \(h(x,y) = axy\) leads to the R\'enyi-entropy \(S_R=\frac{1}{q-1}\ln\int f^q\), with \(s=\frac{1}{1-q}\frac{f^q}{q}\), \(q=1-aT\), and \(X(E)=\frac1a \ln(aE)\). The distribution function is given by \(f(E)=\frac1Z (aE)^{-1/(aT)}\).

	The energy composition rule
	\begin{equation}
		h(x,y) = x+y+axy   \label{eqTsRule}
	\end{equation}
	has the solution 
	\begin{equation}
		X(E)=\frac1a \ln(1+aE),   \label{eqTsX}
	\end{equation}
	and leads to the cut power-law stationary distribution
	\begin{equation}
		f(E)=\frac1Z(1+aE)^{-1/aT}. 
	\end{equation}
	(cf.\ for \(q=1-aT\) the \(q-\)Tsallis distribution emerges \cite{TSBGK})
	
	This form is the next to leading order expansion of a general associative rule for low energy: By Taylor expanding eq.\ \eqref{Xadd} for low \(x\), \(y\) and \(h\) values and requiring \(h(x,0)=x\), \(h(0,y)=y\) one arrives at
	\begin{equation}
		h(x,y)=x+y-\frac{X''(0)}{X'(0)}xy+\dots .
	\end{equation}
	We conclude that the Tsallis-type composition rule with \(a=-X''(0)/X'(0)\) is generic for leading order non-extensive effects at low energy.


\section{Non-extensive parton cascade}

In order to investigate the equilibration of non-extensive systems we start with two subsystems, equilibrated separately. In order to prepare these systems the non-extensive Boltzmann equation is solved numerically in a parton cascade simulation \cite{TSBPGPRL}. We use several different initial momentum distributions, and then make random binary collisions between randomly chosen pairs of particles. By doing so we apply the rules
	\begin{equation}
		X\left(E_1\right) + X\left(E_2\right) = X\left(E_3\right) + X\left(E_4\right),  \label{bcr1}
	\end{equation}
	\begin{equation}
		\vec p_1 + \vec p_2 = \vec p_3 +\vec p_4.  \label{bcr2}
	\end{equation}
In each step of the simulation we select two particles to collide. Then we find the value for the new momentum of the first particle (\(\vec p_3\)) satisfying the above constraints but otherwise random. Then applying eq.\ \eqref{bcr2} we calculate the momentum of the second outgoing particle (\(\vec p_4\)). 
In these particular simulations, this paper reports about, we use the free dispersion relation for massless particles (\(E_i(\vec p_i) = \left|\vec p_i\right|\)). We proceed with the next collision repeating the above steps with a new randomly chosen pair of particles.
A typical simulation includes \(10^6-10^7\) collisions among \(10^5-10^6\) particles. After \(3-5\) collisions per particle on the average, the one-particle distribution approaches its stationary form sufficiently.

The following quantities are conserved during the simulation:
	\begin{equation}
		X\left(E_{tot}\right) = \sum_{i=1}^N X\left(E_i\right), \;\;\;\;
		\vec P = \sum_{i=1}^N \vec p_i, \;\;\;\;
		N = \sum_{i=1}^N 1.
	\end{equation}

We use eq.\ \eqref{eqTsRule} for the energy composition rule, where \(a\) is called the non-extensivity parameter. Our model reconstructs the traditional Boltzmann-Gibbs thermodynamics in the limit of \(a=0\).
More details can be found in \cite{TSBPGPRL}.

We perform simulations on different systems with particle numbers \(N_1\) and \(N_2\), total (quasi-)energies \(X(E_1)\) and \(X(E_2)\) and non-extensivity parameters \(a_1\) and \(a_2\). We evolve these systems until they reach their stationary states. The unified system is taken as an initial state with \(N=N_1+N_2\) particles.

Now we have three non-extensivity parameters \(a_1\), \(a_2\) and \(a_{12}\), describing the three possible interaction types in the composed system. \(a_1\) is the parameter of the collisions among particles in the first subsystem, similarly \(a_2\) corresponds to the second subsystem, and finally \(a_{12}\) is the non-extensivity parameter in a collision between a particle from the first subsystem and a particle from the second subsystem.
We investigate cases with common and different \(a\) parameters.


\section{Results}

\begin{figure}[h]
	{
		\begin{tabular}{cc}
			a\resizebox{0.48\textwidth}{!}{\rotatebox{-90}{\includegraphics{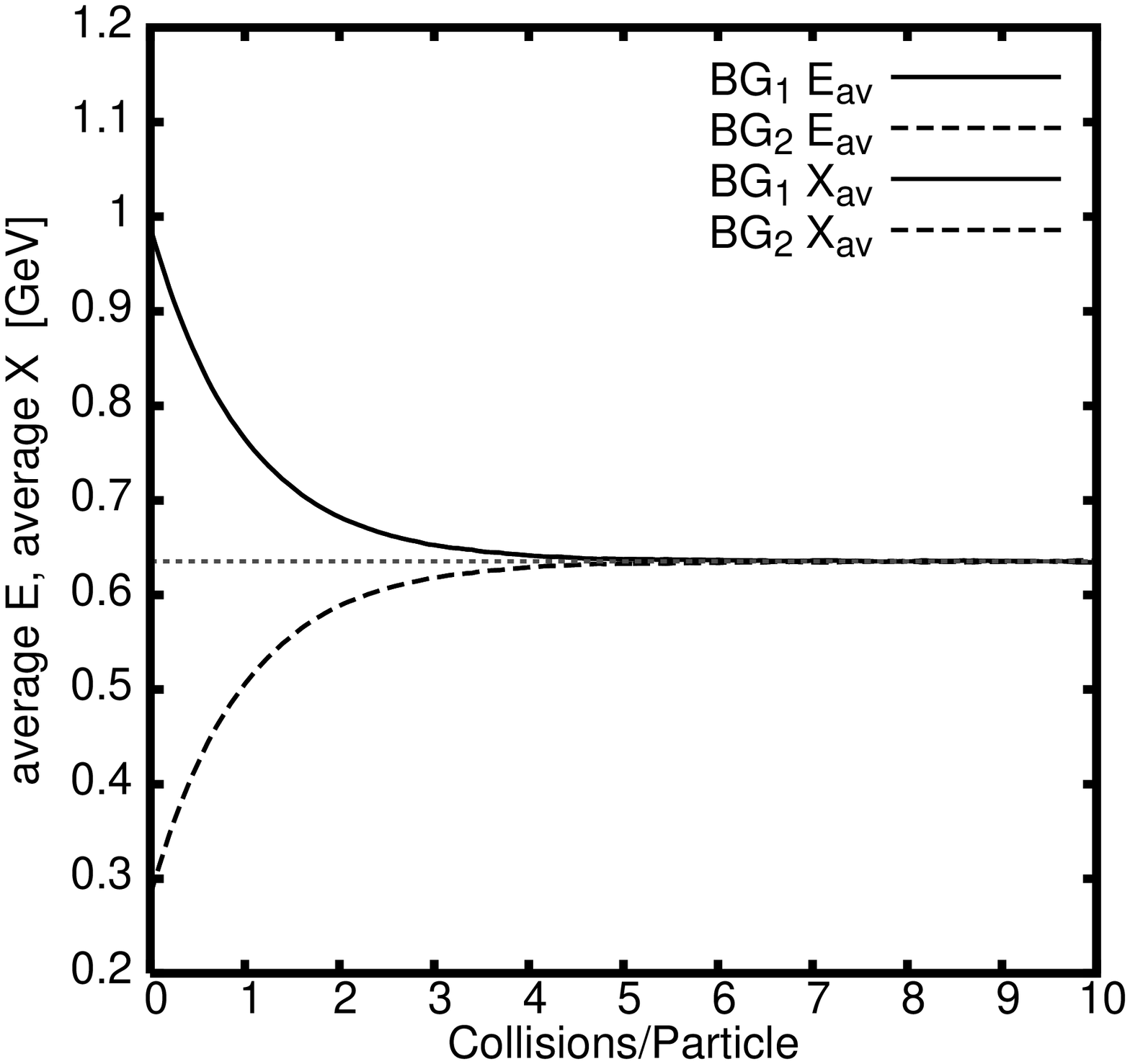}}} & b\resizebox{0.48\textwidth}{!}{\rotatebox{-90}{\includegraphics{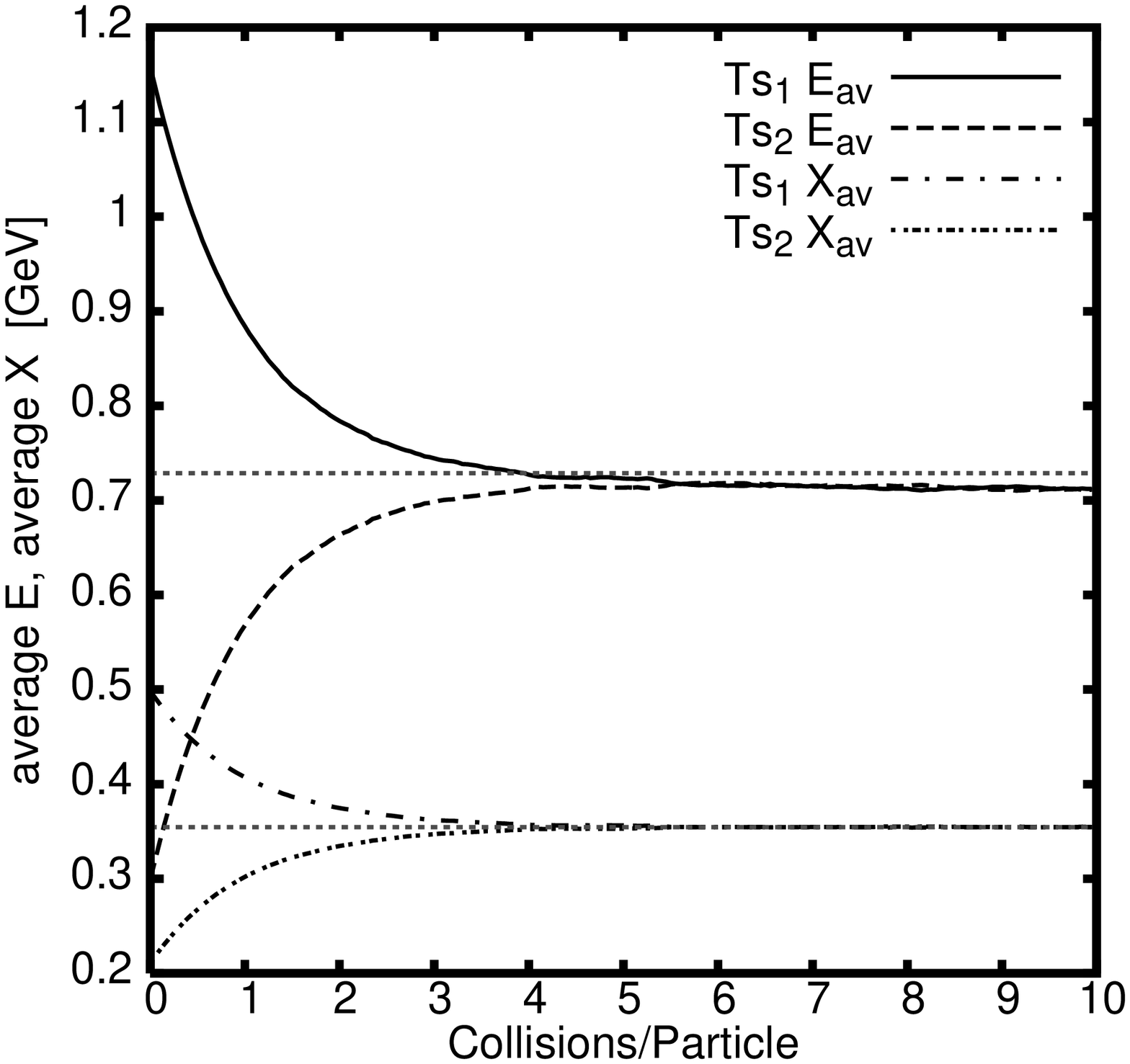}}} 
		\end{tabular}
	}
	\caption{\(a\), Equilibration of two Boltzmann-Gibbs systems (\(a_1=a_2=a_{12}=0\)) \label{fig:XBG};
			 \(b\), Equilibration of two Tsallis-type non-extensive systems (\(a_1=a_2=a_{12}=2\)) \label{fig:XTS}
			}
\end{figure}

\begin{figure}[h]
	{
		\begin{tabular}{cc}
			a\resizebox{0.48\textwidth}{!}{\rotatebox{-90}{\includegraphics{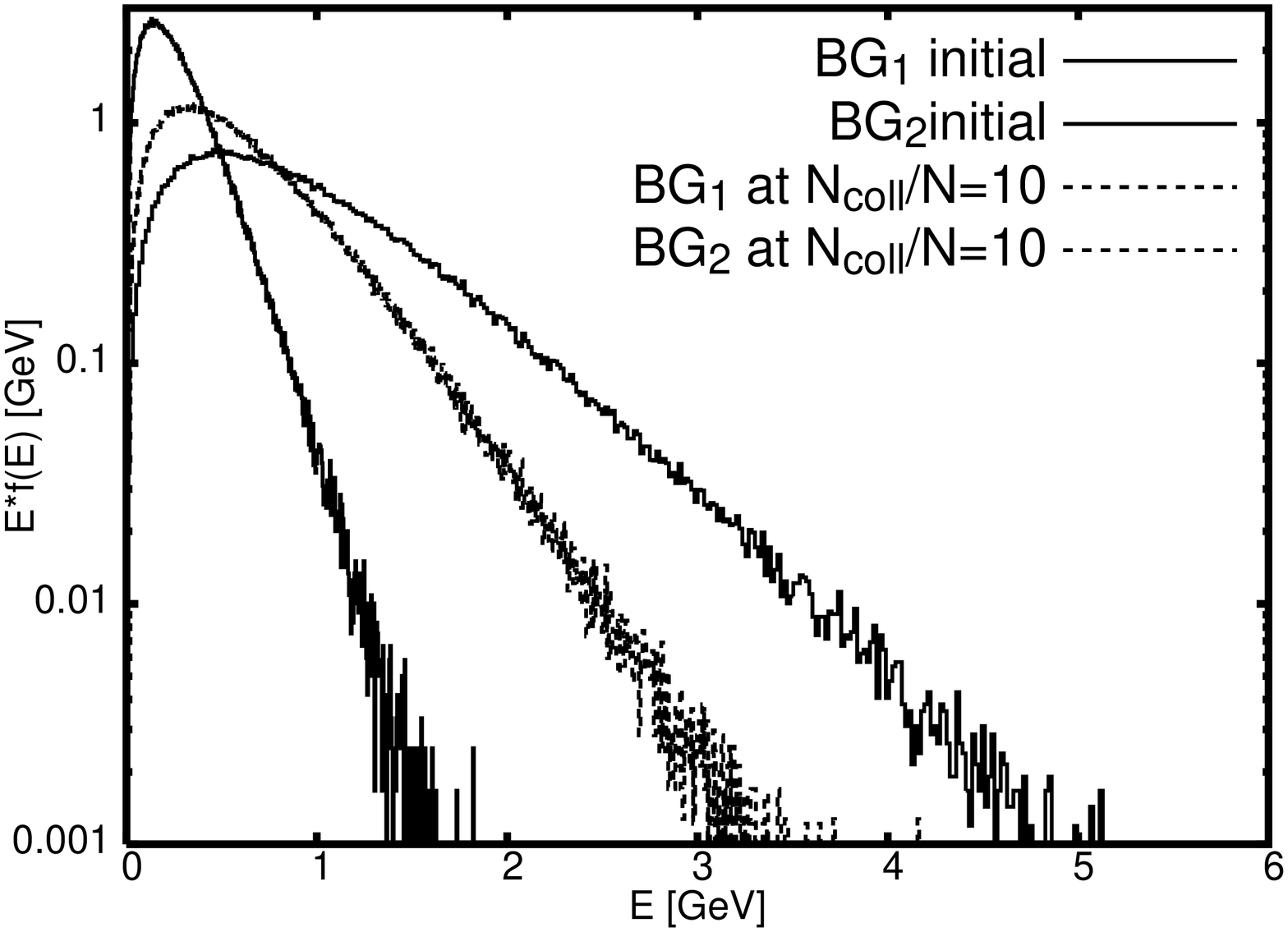}}} & b\resizebox{0.48\textwidth}{!}{\rotatebox{-90}{\includegraphics{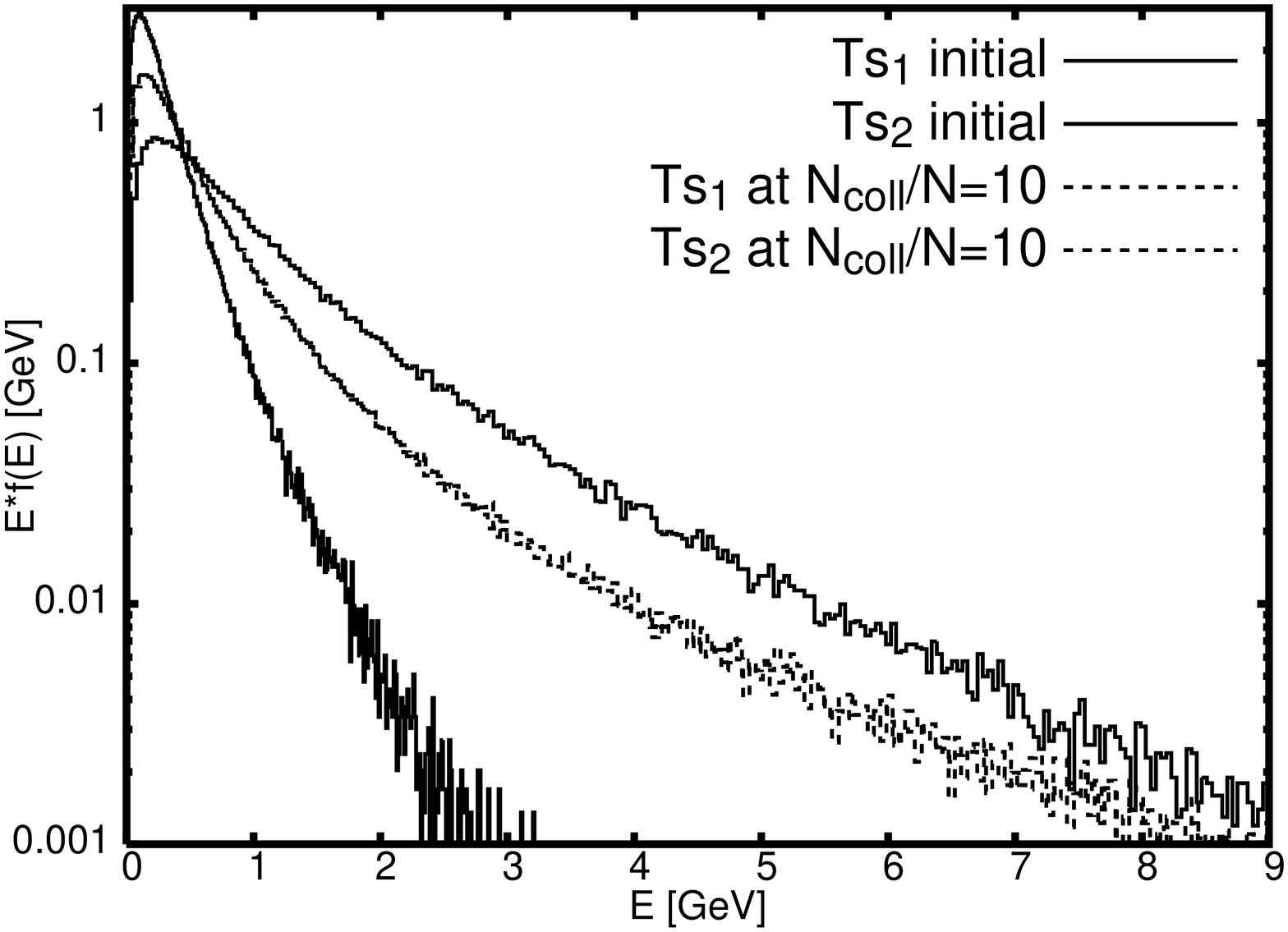}}}
		\end{tabular}
	}
	\caption{\(a\), Equilibration of two Boltzmann-Gibbs systems (\(a_1=a_2=a_{12}=0\)) \label{fig:BG};
			 \(b\), Equilibration of two Tsallis-type non-extensive systems (\(a_1=a_2=a_{12}=2\)) \label{fig:TS}
			}
\end{figure}

\begin{figure}[h]
	{
		\begin{tabular}{cc}
			a\resizebox{0.48\textwidth}{!}{\rotatebox{-90}{\includegraphics{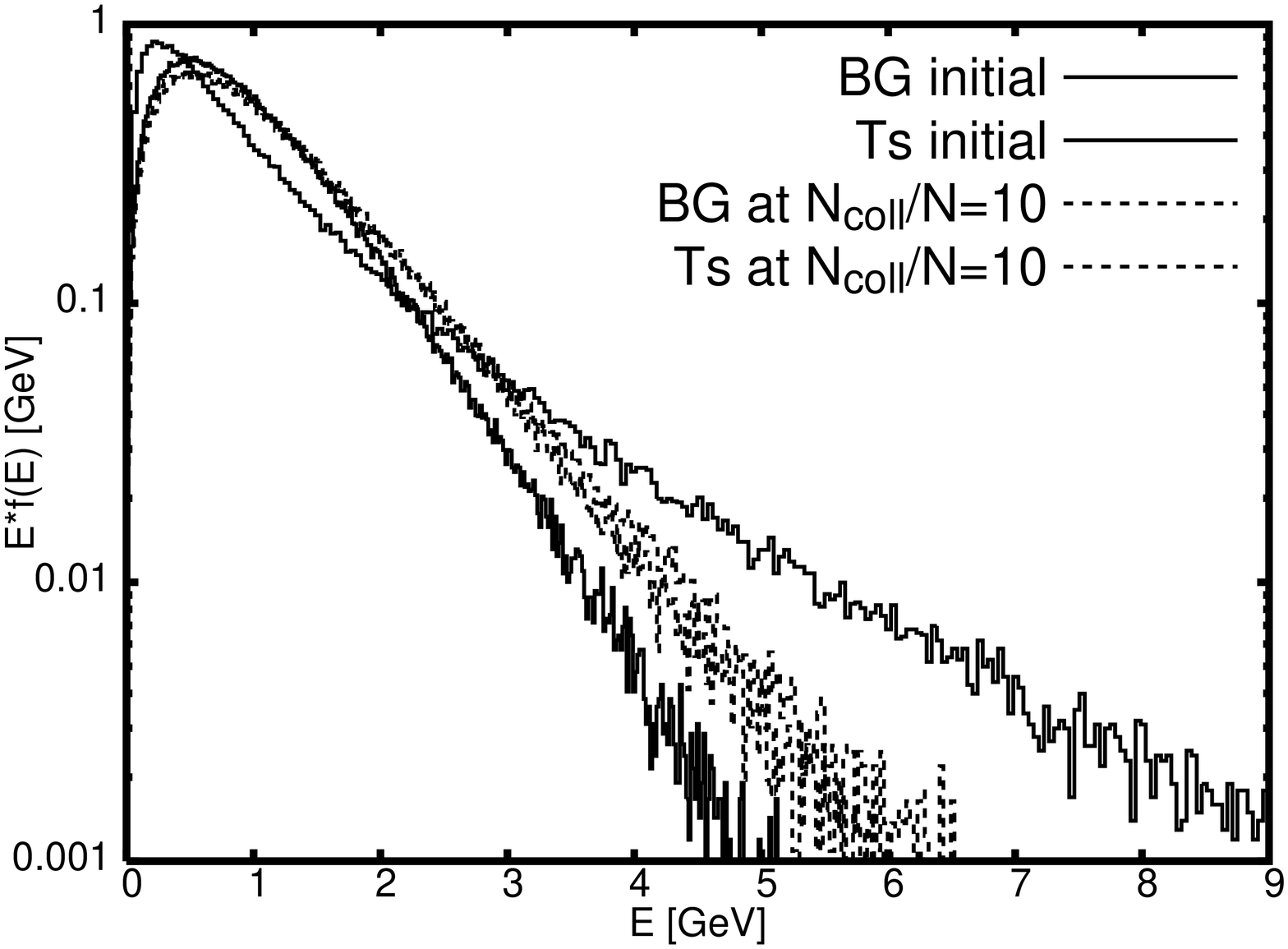}}} & b\resizebox{0.48\textwidth}{!}{\rotatebox{-90}{\includegraphics{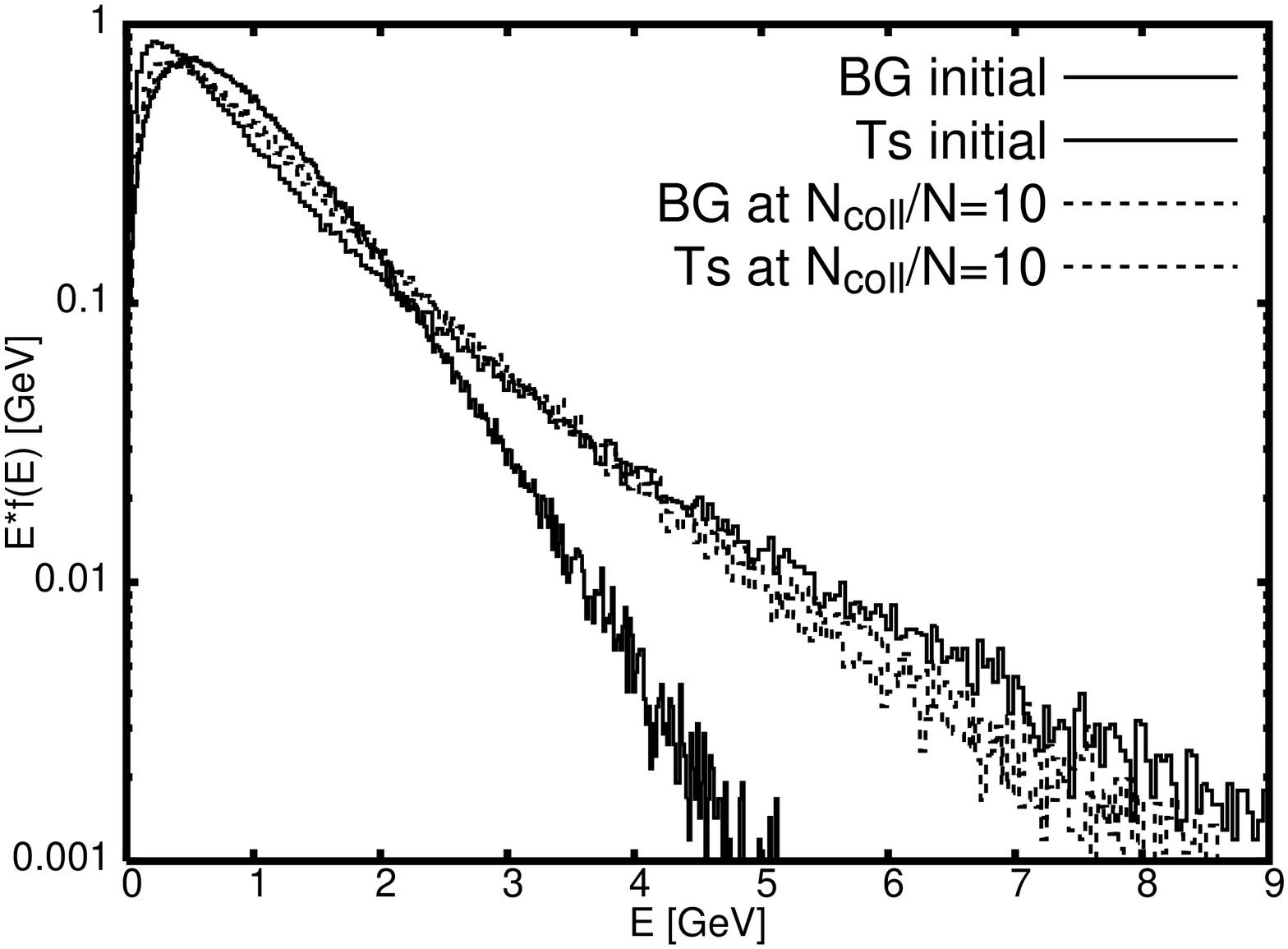}}}
		\end{tabular}
	}
	\caption{\(a\), Equilibration of a Boltzmann-Gibbs and a Tsallis-type non-extensive system (\(a_1=0\), \(a_2=2\) and \(a_{12}=0\)) \label{fig:BGTS0};
			 \(b\), Equilibration of a Boltzmann-Gibbs and a Tsallis-type non-extensive system (\(a_1=0\), \(a_2=2\) and \(a_{12}=2\)) \label{fig:BGTS2}
			}
\end{figure}

Our results show that the subsystems equilibrate, in the final state of the composed system they have a common stationary distribution.

We present examples with different initial conditions. In all of these simulations the particle numbers are the same  for each subsystem, \(N_1=N_2=250\,000\). The number of collisions in one simulation is \(N_{coll}=5\,000\,000\), so \(N_{coll}/(N_1+N_2)=10\) collisions happen per particle. Figure \ref{fig:XBG} shows the average energy and quasi-energy versus the number of collisions per particle. On each part of the figures \ref{fig:BG} and \ref{fig:BGTS0} four curves show the initial (continuous line) and final (dashed line) energy distributions of subsystems \(1\) and \(2\). It is hard to distinguish the final states of the subsystems, because the corresponding distributions are very near to each other. Within numerical uncertainities they have a common stationary energy distribution.

Figures \ref{fig:XBG}\(a\) and \ref{fig:BG}\(a\) show the equilibration of two Boltzmann-Gibbs systems, where \(a_1\), \(a_2\) and \(a_{12}\) are equal to \(0\). This simulation was done for test purposes. As expected, in the final state the two subsystems have a common stationary energy distribution with a common temperature. The energy content per particle of the composed system became the arithmetic mean of the respective initial values. On figure \ref{fig:XBG}\(a\) the curves for average \(X_{av}=\sum_iX(E_i)/N\) coincides with the curve for \(E_{av}=\sum_iE_i/N\), which comes from the definition of the quasi-particle energy (cf.\ eq.\ \eqref{eqTsX} with \(a=0\)).

The equilibration of two Tsallis-type non-extensive systems with \(a_1=a_2=a_{12}=2\) is shown on figures \ref{fig:XTS}\(b\) and \ref{fig:TS}\(b\). The initial quasi-energy contents are different but at the end this difference disappears. \(X_{av}\) becomes the arithmetic mean in the composed system. This is illustrated by the constant for \((X_{1,av}^{initial}+X_{2,av}^{initial})/2\). The final average kinetic energy per particle, \(E_{1,av}^{final}=E_{2,av}^{final}\not=(E_{1,av}^{initial}+E_{2,av}^{initial})/2\), is slightly below the average of the initial energies.

On the figures \ref{fig:BGTS0}\(a\) and \ref{fig:BGTS2}\(b\) we show the equilibration between two systems with different non-extensivity parameters: A Boltzmann-Gibbs system (\(a=0\)) and another one with \(a=2\). In \ref{fig:BGTS0}\(a\) the non-extensivity parameter for a collision between particles coming from different systems is taken to be \(a_{12}=0\). In \ref{fig:BGTS2}\(b\) this parameter is given by \(a_{12}=2\).
In these cases neither \(X_{av}\) nor \(E_{av}\) approaches to the arithmetic mean of the initial \(X_{av}\) or \(E_{av}\) values (actually \(X_{av}\) cannot be defined for the composed system).
The final common stationary energy distributions are between the initial energy distributions of the original subsystems (with \(a_1=0\) and \(a_2=2\)); in case \(a\) (\(a_{12}=0\)) it is closer to the Boltzmann-Gibbs system, in case \(b\) (\(a_{12}=2\)) it is closer to the original non-extensive system.


\section{Conclusions}
We studied the equilibration between two non-extensive systems in the framework of a relativistic parton cascade model, NEBE. We have demonstrated earlier that these systems have non-exponential equilibrium distributions in terms of the free particle kinetic energy, which is dependent on the chosen type of non-extensivity (or equivalently on the details of energy sharing in the two-particle collisions). The composition of different non-extensive systems resulted in a common equilibrium distribution.

We have pointed out that the non-extensive energy addition rule eq.\ \eqref{eqTsRule}, leading to a cut power-law distribution in equilibrium, is the next to leading order low-energy expansion of a general associative rule.

We coupled two subsystems after a separate equilibration both by additive and super-additive rules. The subsystems tend to achieve a common distribution. This process coincides with the equilibration of the temperatures in the case of equal non-extensivity parameters. In the case of different non-extensivity parameters, however, neither the average of original one-particle energies nor the average of the quasi-particle energies is observed. In this case the associativity of the energy composition rule is violated, so actually a quasi-particle energy can not be defined.


\section*{Acknowledgments}
This work has been supported by the Hungarian National Science Fund OTKA (T049466) and the Deutsche Forschungsgemeinschaft.


\end{document}